\documentstyle[12pt]{article}

\font\ss=msbm10

\newcommand{\be}{\begin{equation}}
\newcommand{\ee}{\end{equation}}
\newcommand{\bean}{\begin{eqnarray*}}
\newcommand{\eean}{\end{eqnarray*}}
\newcommand{\bea}{\begin{eqnarray}}
\newcommand{\eea}{\end{eqnarray}}
\newcommand{\RR}{\hbox{\ss R}}
\newcommand{\diag}{{\rm diag}}
\newcommand{\sign}{{\rm sgn}}
\newcommand{\tr}{{\rm tr}}

\newcommand{\sq}{\sqrt{|g|}}

\def\dil#1{e^{-\alpha_{#1}\cdot\phi}}

\author{M. G. Ivanov\thanks{e-mail: mgi@mi.ras.ru}}
\title{Intersecting delocalized p-branes}
\date{July 24, 2001}

\begin{document}

\maketitle

\abstract{
  A model considered in the paper generalizes supergravity type model
 to the case of delocalized membrane sources.
  A generalization of intersecting p-brane solution with delocalized
 membranes is presented.
}

\section{Introduction}

  Membrane theories \cite{GSW,BN,VV,Duff} consider membranes as mapping $x$
 from manifold ${\bf V}$ of dimension $n$ to (pseudo)Riemannian manifold
 ${\bf M}$ of dimension $D>n$.
  The manifold ${\bf M}$ is interpreted as space-time (``bulk'').
  The image $x({\bf V})$ is interpreted as membrane.
  Membrane appears in bulk equation in the form of singular sources localized
 at $x({\bf V})$.
  A procedure of regularization can be considered as replacement of
 common membrane, which has zero thickness by some sort of thick membrane,
 which corresponds to continuous distribution of infinitely light membranes.

  A covariant method of regularization of closed membrane theories
 (``$({\bf M\to F})$-approach'') was presented in the papers
 \cite{MGI-DAN,_h_mem}.
  $({\bf M\to F})$-approach replaces mapping $x:{\bf V\to M}$
 by mapping $\varphi:{\bf M\to F}$, where ${\bf F}$ is a manifold
 of dimension $D-n$.
  If for some point $\phi\in{\bf F}$ inverse image $\varphi^{-1}(\phi)$
 is submanifold of dimension $n$, then it has to be considered
 as infinitely light membrane.
  The similar ideas were suggested before in papers \cite{hos1,morris1,bf1}
 (see also references in paper \cite{_h_mem}).

  The generalization of p-brane solution with delocalized membranes
 was also presented in the papers \cite{MGI-DAN,_h_mem}.
  A natural generalization of extremal intersecting p-brane
 solution with delocalized membranes is constructed in the present paper
 (on intersecting $p$-brane solutions see
 \cite{AIR,IM} and references in these papers).

\section{Notation}

  $D$-dimensional space-time, i.e. smooth (pseudo)Riemannian
 manifold ${\bf M}$ with metric 
 $ds^2=g_{MN}dX^MdX^N$, $M,N,\dots=0,\dots,D-1$
 is considered.
  In further calculations ${\bf M}$ is considered to be
 a finite region in $\RR^D$ with smooth boundary $\partial{\bf M}$.

  For differential forms $A$ and $B$ with components
 $A_{M_1\dots M_q}$ and $B_{N_1\dots N_p}$
 the following tensor is introduced
$$
 (A,B)_{M_{k+1}\dots M_q~N_{k+1}\dots N_p}^{(k)}=\frac{1}{k!}~
         g^{M_1N_1}\dots g^{M_kN_k}
         A_{M_1\dots M_q}B_{N_1\dots N_p}.
$$
  Index $(k)$ indicates the number of indices to contract.
  If it can not lead to ambiguity, $(k)$ is skipped.

  For differential form of power $q$ it is
 convenient to introduce two norms
 $\|A\|^2_\Sigma=\Sigma(A,A)^{(q)}$,
 where $\Sigma=\pm1$,
 and Hodge duality operation $*A=(\Omega,A)^{(q)}$,
 where $\Omega=\sq~d^DX$ is form of volume.
  Here and below $g=\det(g_{MN})$, $\sigma=\sign(g)$.

\section{Action and equations of motion}

  Let us consider the action of the following form
\bea
  &&S=\int\limits_{\bf M}d^DX\sq
   \left(
     \frac{R}{2\kappa^2}
    -\frac12\|\partial\phi\|^2
      -\sum_I
         \frac12\beta_Ie^{-\alpha_I\cdot\phi}\|F^{(I)}\|^2
   \right.
\nonumber
\\
     &&\left.
        +\sum_a\lambda_{(I,a)} e^{\frac12\alpha_I\cdot\phi}
         \|J^{(I,a)}\|_{\Sigma_{(I,a)}\sigma}
        +s_{(I,a)}\frac{\lambda_{(I,a)}}{h_{(I,a)}}(*J^{(I,a)},A^{(I)})
   \right).
\label{S-pb}
\eea

  Here
 $\phi=(\phi^1,\dots,\phi^{n_\phi})$ is a set of scalar fields (dilatons),
 $\alpha_I\cdot\phi=\sum_{i=1}^{n_\phi}\alpha_I^i\phi^i$,
 $F^{(I)}=dA^{(I)}$ are exact $(q_I+1)$-forms,
 $J^{(I,a)}=d\varphi^1_{(I,a)}
  \wedge\dots\wedge d\varphi^{D-q_I}_{(I,a)}$,
 $\Sigma_{(I,a)}=\pm1$, $s_{(I,a)}=\pm1$,
 $\alpha_I^i$, $\lambda_{(I,a)}$ and $h_{(I,a)}$ are real constants.

  By variation of action (\ref{S-pb}) over metric
 $g_{MN}$, scalar fields $\phi$, $q_I$-forms $A^{(I)}$
 and membrane potentials $\varphi^\alpha_{(I,a)}$ one
 can find equations of motion.
  Einstein equations, found by variation of metric have the form
\be
  R_{MN}-\frac12Rg_{MN}=\kappa^2 T_{MN},
\label{Ein1pb}
\ee
 where energy-momentum tensor is defined by the following formulae
\bea
  &&T_{MN}=T^{(\phi)}_{MN}
   +\sum_I(T^{(I)}_{MN}+\sum_a T^{{(I,a)}}_{MN}),
\label{emt0pb}\\
  &&T^{(\phi)}_{MN}=
      \partial_M\phi\cdot\partial_N\phi-\frac12\|\partial\phi\|^2g_{MN},
\label{emt1pb}\\
  &&T^{(I)}_{MN}=\beta_I\dil{I}
    \left(
      (F^{(I)},F^{(I)})_{MN}-\frac12\|F^{(I)}\|^2g_{MN}
    \right),
\label{emt2pb}\\
  &&T^{(I,a)}_{MN}=-\lambda_{(I,a)}e^{\frac12\alpha_I\cdot\phi}
     \|J^{(I,a)}\|_{\Sigma_{(I,a)}\sigma}
     \left(g_{MN}-\frac{(J^{(I,a)},J^{(I,a)})_{MN}}{\|J^{(I,a)}\|^2}\right).
\label{emt3pb}
\eea
  Equations, found by variation of forms $A^{(I)}$, have the form
\be
  \beta\delta\left(\dil{I}F^{(I)}\right)
  =\sigma(-1)^{q_I+1}Q^{(I)},
\label{Max1pb}
\ee
 where the source $Q^{(I)}$ defined by
\bea
  Q^{(I)}=-\sum_a
    s_{(I,a)}\frac{\lambda_{(I,a)}}{h_{(I,a)}}~*J^{(I,a)}.
\label{Q1pb}
\eea
  Equations, found by variation of scalar fields, have the form
\be
  \Box\phi=-\sum_{I}(Q_{\phi I}+Q_{\phi Is}),
\label{phi1pb}
\ee
 where the sources are defined by
\bea
  Q_{\phi I}&=&\frac12\alpha_I\beta_I\dil{I} \|F^{(I)}\|^2,
\label{Qphi1pb}\\
  Q_{\phi Is}&=&-\frac{\alpha_I}2
    e^{\frac12\alpha_I\cdot\phi}
    \sum_a\lambda_{(I,a)}
    \|J^{(I,a)}\|_{\Sigma_{(I,a)}\sigma}.
\label{Qphi2pb}
\eea
  Equations, found by variation of membrane potentials, have the form
\be
  \left(\delta
  \left[
    \Sigma_{(I,a)}\sigma
    \frac{\dil{I}J^{(I,a)}}{\|J^{(I,a)}\|_{\Sigma_{(I,a)}\sigma}}
   +(-1)^{q_I(D-q_I)}s_{(I,a)}
    \frac{*A^{(I,a)}}{h_{(I,a)}}
  \right],
  d\varphi^{\alpha_1}_{(I,a)}\wedge\dots\wedge
  d\varphi^{\alpha_{D-q_I-1}}_{(I,a)}\right)=0.
\label{varphi1pb}
\ee

\section{Delocalized p-brane solutions}

  The goal of the paper is to present special solutions of equations
 (\ref{Ein1pb}), (\ref{Max1pb}), (\ref{phi1pb}) and (\ref{varphi1pb}).

  One can introduce a set of orthonormal projectors
 $P^{{(I,a)}~M}_N$,
 $\bar P^{{(I,a)}~M}_N=\delta^M_N-P^{{(I,a)}~M}_N$
 numerated by index ${(I,a)}$.
  In the given coordinate system
\be
  \tr\bar P^{(I,a)}=\bar P^{{(I,a)}~M}_M=q_I;~~~~
  P^{{(I,a)}~M}_N=0,~~M\not=N,
\ee
  i.e. $\bar P^{{(I,a)}M}_N=\Delta_{{(I,a)}M}\delta_{MN}$
 (there is no summeation over $M$), $\Delta_{{(I,a)}M}\in\{0,1\}$.

  Let us define fields by the following equations
\bea
\label{A0pb}
  A^{(I)}&=&\sum_a\frac{h_{(I,a)}}{H_{(I,a)}}~\omega^{(I,a)},
\eea
 where
\be
  \omega^{(I,a)}= e^{W_{(I,a)}}~
  \bigwedge_{\Delta_{{(I,a)}M}=1}dX^M,
\ee
 $h_{(I,a)}$ are constants,
 $H_{(I,a)}$ are smooth positive functions
 ({\em parametrizing functions}),
 $W_{(I,a)}$ are smooth functions, which satisfy conditions
 $(P^{(I,a)}\partial)_MW_{(I,a)}=0$.

\be
  ds^2=\left(\prod_{I,a}
   H_{(I,a)}^{2\kappa^2\Sigma_{(I,a)}\beta_I h^2_{(I,a)} 
    \left(
       \Delta_{{(I,a)}M}
         -\frac{q_I}{D-2}\right)}\right)\eta_{MN}dX^MdX^N,
\label{ds0pb}
\ee
 where $\eta_{MN}=\diag(\pm1,\dots,\pm1)$.

\be
  \phi=\frac12~\sum_{I,a}\Sigma_{(I,a)}\beta_I h^2_{(I,a)}
          ~\alpha_I~\ln H_{(I,a)}.
\label{phi0pb}
\ee

\be
  J^{(I,a)}={\cal J}_{(I,a)}~*\omega^{(I,a)}.
\label{J0pb}
\ee

  {\Large Theorem on $p$-brane solutions with sources.}
  {\sl If fields defined by equations
   (\ref{A0pb}), (\ref{ds0pb}), (\ref{phi0pb}), (\ref{J0pb}),
   satisfy the following conditions}
\bea
  &&|{\cal J}_{(I,a)}|=\frac{\Sigma_{(I,a)}\beta_I h^2_{(I,a)}}
                      {\lambda_{(I,a)}}~
  e^{-\frac12\alpha_I\cdot\phi}
      ~\frac{\Box H_{(I,a)}}{H^2_{(I,a)}},
\\
  &&\sign\|\omega^{(I,a)}\|^2=\Sigma_{(I,a)},
\\
  &&(\bar P^{(I,a)}\partial)_MH_{(I,a)}=0,
\eea
\be
  (P^{(I,a)}\partial)_M H_{(I,a)}~
  (\bar P^{(I,a)}\partial)_N
  \left(
    \kappa^2\sum_{J,b}\Sigma_{(J,b)}\beta_{J}h^2_{(J,b)}
    \ln H_{(J,b)}~2
      \Delta_{{(I,a)}M}
     +W_{(I,a)}
  \right)=0,
\ee
\be
  \ln H_{(I,a)}
  =\kappa^2\sum_{J,b}\Sigma_{(J,b)}\beta_{J}h^2_{(J,b)}\ln H_{(J,b)}
   \left[
     \frac{q_Iq_J}{D-2}-\tr(\bar P^{(I,a)}\bar P^{(J,b)})
    -\frac{\alpha_I\cdot\alpha_J}{4\kappa^2}
   \right]+W_{(I,a)},
\ee
\bea
  &&\tr(\bar P^{(I,b)}\bar P^{(I,c)})\not=q_I-1,
\\
  && s_{(I,a)}=-\Sigma_{(I,a)}\sigma(-1)^{q_I(D-q_I)}
               \sign{\cal J}_{(I,a)},
\eea
  {\sl Then fields
   (\ref{A0pb}), (\ref{ds0pb}), (\ref{phi0pb}), (\ref{J0pb})
   satisfy equations of motion
   (\ref{Ein1pb}), (\ref{Max1pb}), (\ref{phi1pb}), (\ref{varphi1pb}).}

  If $\Box H_{(I,a)}=0$ for all ${(I,a)}$, then
 membrane fields $J^{(I,a)}$ vanish and the solution
 is a regular intersecting extremal p-brane solution
 (see \cite{AIR,IM} and references in these papers for examples).
  Vice-versa, one can introduce non-trivial membranes fields
 to intersecting extremal electric-type p-brane solution by setting
 $\Box H_{(I,a)}\not=0$.

\section{Examples}
  As a simplest example, let us consider solution of
 Einstein-Maxwell equations in the presence
 dust cloud (with zero pressure)
 with charge density equal to mass density.

  The action has the following form
$$
  S=\int d^4X\sqrt{-g}
    \left(
      \frac{R}2-\frac{\|F\|^2}{2}
     -\|J\|-\frac1{\sqrt{2}}~(*J,A)
    \right),
$$
 where $F=dA$ is electromagnetic field,
 $A$ is 4-vector potential,
 $J=d\varphi^1\wedge d\varphi^2\wedge d\varphi^3$.

  By variation over fields $g_{MN}$, $A_M$, $\varphi^\alpha$
 one finds the following equations of motion
$$
  R_{MN}-\frac12Rg_{MN}=(F,F)_{MN}+\frac{(J,J)_{MN}}{\|J\|}
   -\left(\frac12\|F\|^2+\|J\|\right)g_{MN},
$$
$$
  \left(\delta\left(\frac{J}{\|J\|}-\frac{*A}{\sqrt2}\right),
    d\varphi^\alpha\wedge d\varphi^\beta\right)=0,
~~
  \delta F=-\frac{*J}{\sqrt2}.
$$
  The fields defined by the following equations solve the
 equations of motion
$$
  ds^2=-H^{-2}dX^0dX^0
       +H^2\delta_{\alpha\beta}dX^\alpha dX^\beta,
$$
$$
  A=\frac{\sqrt2}H~dX^0,
~~
  J=-2\triangle H~dX^1\wedge dX^2\wedge dX^3.
$$
  Here $H$ is smooth positive function
 of space coordinates $X^\alpha$, $\alpha,\beta=1,2,3$,
 $\triangle=\partial_1^2+\partial_2^2+\partial_3^2$.
  Calculating $\|J\|$ one has to choose a branch of square root,
 which gives $\|J\|\triangle H<0$.

  The following example describes intersecting delocalized
 2-branes in 11d supergravity.

  The action has the following form
$$
  S=\int d^{11}X\sqrt{-g}
    \left(
      \frac{R}2-\frac{\|F\|^2}{2}
     -\|J^{(1)}\|-\|J^{(2)}\|+\sqrt2~(*J^{(1)}+*J^{(2)},A)
    \right),
$$
 where $F=dA$,
 $A$ is form of power 3,
 $J^{(a)}=d\varphi^1_a\wedge\dots\wedge d\varphi^8_a$,
 $a=1,2$.

  By variation over fields $g_{MN}$, $A_{MNK}$, $\varphi^\alpha_a$
 one finds the following equations of motion
$$
  R_{MN}-\frac12Rg_{MN}=(F,F)_{MN}+\frac{(J^{(1)},J^{(1)})_{MN}}{\|J^{(1)}\|}
   +\frac{(J^{(2)},J^{(2)})_{MN}}{\|J^{(2)}\|}
   +\left(\frac12\|F\|^2+\|J^{(1)}\|+\|J^{(2)}\|\right)g_{MN},
$$
$$
  \left(\delta\left(\frac{J^{(a)}}{\|J^{(a)}\|}
             -\sqrt2~*A\right),
  d\varphi^{\alpha_1}_a\wedge\dots\wedge
  d\varphi^{\alpha_7}_a\right)=0,
~~
  \delta F=-\sqrt2(*J^{(1)}+*J^{(2)}).
$$
  The fields defined by the following equations solve the
 equations of motion
$$
  ds^2=-H_1^{-2/3}H_2^{-2/3}dX^0dX^0
       +H_1^{-2/3}H_2^{1/3}(dX^1dX^1+dX^2dX^2)
$$
$$
       +H_1^{1/3}H_2^{-2/3}(dX^3dX^3+dX^4dX^4)
       +H_1^{1/3}H_2^{1/3}\sum_{\alpha=5}^{10}dX^\alpha dX^\alpha,
$$
$$
  A=\frac1{\sqrt{2}H_1}~dX^0\wedge dX^1\wedge dX^2
   +\frac1{\sqrt{2}H_2}~dX^0\wedge dX^3\wedge dX^4,
$$
$$
  J^{(1)}=-\frac12\triangle H_1~dX^3\wedge\dots\wedge dX^{10},
$$
$$
  J^{(2)}=-\frac12\triangle H_2~dX^1\wedge dX^2\wedge
           dX^5\wedge dX^6\wedge\dots\wedge dX^{10}.
$$
  Here $H_a$ are smooth positive function
 of space coordinates $X^\alpha$, $\alpha,\beta=5,6,7,8,9,10$,
 $\triangle=\sum_{\alpha=5}^{10}\partial_\alpha^2$.
  Calculating $\|J^{(a)}\|$ one has to choose a branch of square root,
 which gives $\|J^{(a)}\|\triangle H_a<0$.

  The author is grateful to I.V. Volovich and B. Dragovich.
  The work was partially supported by grant RFFI 99-01-00866.


\end{document}